\documentclass[a4paper,notitlepage,12pt]{article}

\begin{document}
\title{
Analysis of correlation structures in the $Synechocystis$ PCC6803 genome}
\author{Zuo-Bing Wu\footnotemark[1]\\
State Key Laboratory of Nonlinear Mechanics,\\ Institute of
Mechanics,\\ Chinese Academy of Sciences, Beijing 100190, China}
 \maketitle

\footnotetext[1]{Correspondence to. Tel.: 86-10-82543955. Email
address: wuzb@lnm.imech.ac.cn}

\newpage
\begin{abstract}
Transfer of nucleotide strings in the $Synechocystis$
sp. PCC6803 genome is investigated to exhibit periodic and non-periodic correlation structures
by using the recurrence plot method and the phase space reconstruction technique.
The periodic correlation structures are generated by periodic transfer of
several substrings in long periodic or non-periodic nucleotide strings
embedded in the coding regions of genes.
The non-periodic correlation structures are generated by non-periodic transfer
of several substrings covering or overlapping with the coding regions of genes.
In the periodic and non-periodic transfer,
some gaps divide the long nucleotide strings into the substrings
and prevent their global transfer.
Most of the gaps are either the replacement of one base
or the insertion/reduction of one base.
In the reconstructed phase
space, the points generated from two or three steps for the continuous iterative
transfer via the second maximal distance can be fitted by two lines.
It partly reveals an
intrinsic dynamics in the transfer of nucleotide strings.
Due to the comparison of the relative positions and lengths,
the substrings concerned with the non-periodic correlation structures are almost identical to
the mobile elements annotated in the genome. The mobile elements are thus endowed
with the basic results on the correlation structures.

\end{abstract}

\newpage
\section{Introduction}

The rapid accumulation of complete DNA sequences of many organisms
provides an opportunity to systematically analyze their components, structures
and functions. On the one hand, from the point of view of statistics and geometry,
nontrivial statistical characteristics, such as the long-range correlations, the short-range
correlations and the fractal features or genomic signatures were
determined\cite{Jeffrey,LK,Peng,Karlin,DGVFF,Hao,RRP,SG,MAL,PO}.
Meanwhile, lots of graphical methods such as dot plot, dot matrix and recurrence analysis
to compare the genomes and visualize their similarity were developed\cite{Mount,CTG,CC,FS,KS}.
On the other hand, it was found that the transposable elements as the mobile
DNA sequences can move in the genomes and make many replicas\cite{BEN,FJW,Kazazian}.
Understanding their origin, evolution, and effects on genome structures
and gene functions
is of fundamental importance for biology\cite{Pe,BQ,LS,De}.

The $Synechocystis$ sp. PCC6803 ({\it synecho}) is one of
unicellular cyanobacteria, which presumably are the oldest
organisms capable of oxygenic photosynthesis.
The transformable ability of the $synecho$ facilitates its biotechnological applications\cite{Thiel}.
Since the entire $synecho$ genome was
determined\cite{Kaneko}, a series of studies on its physical and
genetic maps, and functions has been
completed\cite{Kotani,Bhaya,Kucho,TSM}. In particularly, 10-11 base
pair oscillations in the statistical correlation analysis were
found to reflect protein structure and DNA folding\cite{HWT}. The
whole genome offers
 meaningful information for understanding the metabolic network
 and transcriptional organization
of this organism in the bioengineering
application\cite{HL,Fu,Knoop,Mitschke}.

So far, the statistical analysis of the $synecho$
genome has exhibited well global properties of base pairs with a
correlation distance.
 However, correlation properties of nucleotide strings
repeated in the genome are neglected, such as the transfer of
nucleotide strings may happen at many positions of the genome and
generate periodic correlation structures\cite{Wu5}.
Besides the repetition of basic periodic nucleotide strings,
the transfer of non-periodic nucleotide strings
with the same increasing periods
would form the periodic correlation structures.
It inspires to explore more extensive correction of nucleotide strings
and the intrinsic mechanism.
 In this paper, by using the recurrence plot method and
 the phase space reconstruction technique, we
identify transfer of nucleotide strings in the $synecho$ genome
and make a detail analysis of the periodic and non-periodic
correlation structures.

\section{Methods}

For a given genome $s_1s_2 \cdots s_i \cdots s_N$ ($s_i \in
{A,C,G,T}$), the mutual information function is defined\cite{shannon,wtli1990,herzel} as
\begin{equation}
I(m) = \sum_{\xi,\eta=1}^4 p_{\xi\eta}(s_{\xi},s_{\eta}) \log_2 \frac{p_{\xi\eta}(s_{\xi},s_{\eta})}{p_{\xi}(s_{\xi}) p_{\eta}(s_{\eta})},
\end{equation}
where $p_{\xi\eta}$ is the relative frequency of the pair of $s_{\xi}$ and
$s_{\eta}=s_{\xi+m}$ in a distance $m$ and $p_{\xi}$ is the relative frequency of
$s_{\xi}$. The $synecho$ genome denoted as BA000022 is obtained
from the GenBank(ftp.ncbi.nih.gov) and has 3573470 bases.
The symmetrical distribution of four bases along the single strand is
$p_{\xi}$=26.1\%/26.2\% and 23.8\%/23.9\%
for $s_{\xi}$=A/T and C/G. Figure 1 displays the
mutual information function of the genome ($m \leq 1000$) and
reflects correlations with the exponential decay in the long-range scale.
 To investigate correlations in the short-range scale, we concentrate on the
 mutual information function for $m \in [1, 100]$ and redraw the local
blow-up region in Fig. 1. It is evident that the fundamental vibration
frequency is 3 base pairs, which are due to the genetic code.
The correlation analysis of the genome provides the global correlation properties
of the two base pairs ($s_{\xi}$ and $s_{\eta}$) with correlation distance $m$ in the short- and
long-range scales, but correlation
properties of two nucleotide strings with the ending bases $s_{\xi}$ and
$s_{\eta}$ in the genome are neglected.

In what follows, we give a brief presentation of the recurrence
plot method based on the metric representation, which is detailed
in\cite{EKR,Wu1,Wu3}. Firstly, the genome is partitioned into
 $N$ subsequences $\Sigma_k= s_1 s_2 \cdots s_k (1 \le k \le N)$
 and mapped in a metric plane ($\alpha, \beta$).
 The metric mapping ($\alpha_k, \beta_k$) of a subsequence
 is defined as
 \begin{equation}
 \begin{array}{l}
 \alpha_k  = 2\sum_{j=1}^k \mu_{k-j+1} 3^{-j} +3^{-k}= 2\sum_{i=1}^k \mu_i 3^{-(k-i+1)}
 +3^{-k},\\
 \beta_k = 2\sum_{j=1}^k \nu_{k-j+1} 3^{-j} +3^{-k}= 2\sum_{i=1}^k \nu_i 3^{-(k-i+1)} +3^{-k},
 \label{eq1}
 \end{array}
 \end{equation}
 where $\mu_i$ is 0 if $s_i \in \{A,C\}$ or 1 if $s_i \in \{G,T\}$
 and $\nu_i$ is 0 if $s_i \in \{A,T\}$ or 1 if $s_i \in \{C,G\}$.
 The points ($\alpha_k, \beta_k$)
 concentrate in local zones of the metric plane ($[0,1] \times [0,1]$).
The subsequences with the
same ending $l$-nucleotide string labeled by $\Sigma^{l}$
 correspond to points in the zone encoded by the
$l$-nucleotide string. With two subsequences $\Sigma_i \in
\Sigma^{l}$ and $\Sigma_j$ ($j \geq l$), we calculate
the distance between the points $(\alpha_i, \beta_i)$ and $(\alpha_j, \beta_j)$
in the plane. When the distance is not longer than the zone size $\epsilon_l=3^{-l}$,
i.e., $\Sigma_j \in \Sigma^{l}$,
 the point $(i,j)$ is plotted in a recurrence plot plane.
Repeating the above process and shifting forward, we obtain the recurrence
 plot of the genome.
 In comparison with the definition (1),
  it is clear that the mutual information function only corresponds to the recurrence plot with $l=1$.
 The recurrence plot basically depends on the length $N$ of the genome and the zone size $\epsilon_l$.
 At the fixed length $N$,
 the recurrence plot with a small $l$
 is easier to investigate global properties
than that with a large $l$, but to find local
properties such as the transfer of long nucleotide
strings, latter is better.
Although the density of points in the recurrence plot
 decreases monotonically as $l$ increases,
 their distributions in the plane are fixed.
 From the recurrence plot
plane, we calculate the maximal value of $x$ to satisfy
 $\Sigma_{i+x}$, $\Sigma_{j+x} \in \Sigma^l$ $(x=0,1,2,\cdots x_{max})$. The
transferred nucleotide string has the length $L=l+x_{max}$ and is placed at
the positions $(i-l+1, i+x_{max})$ and $(j-l+1, j+x_{max})$, which
implies the transferring distance $d_T=j-i$ starting from the diagonal line in the plane.
 Then, to depict the correlation structures
 in the plane, on the one hand, we propose a correlation intensity
 at a given transferring distance $d_T$
\begin{equation}
\Xi(d_T) = \sum_{i=1}^{N-d_T}
\Theta(\epsilon_{l}-|\Sigma_i-\Sigma_{i+d_T}|), \label{eq4}
\end{equation}
where $\Theta$ is the Heaviside function;
on the other hand, we define an iterative transferring distance $x_k$ of the given nucleotide string
\begin{equation}
\begin{array}{l}
x_{k}=d_T(k)-d_T(k-1),\ \ \ k>1,\\
x_1=d_T(1),
\end{array}
\end{equation}
where $d_T(k)$ is the $k$-th transferring distance of the given nucleotide string
starting from the diagonal line in the plane.
Applying the phase space reconstruction technique\cite{PCFS},
we generate two-dimensional vectors from the one-dimensional iterative transferring distance $x_k$
\begin{equation}
{\bf y}_k=(x_k,x_{k+1}),\ \ \ k=1,2 \cdots.
\end{equation}

\section{Analytical results}

\subsection{Recurrence analysis of coding and non-coding regions}

 By using the above method in Sect. 2, points in the recurrence plot of the genome for $l=15$
 are determined and divided into four parts shown in Fig. 2 due to the transfer of $l$-nucleotide
 strings in/between coding and/or non-coding regions of the genome.
 Figures 2(a) and 2(d) display
 the transfer of $l$-nucleotide strings in the coding and the non-coding regions, respectively.
 One half is excluded due to the mirror symmetry about the diagonal line $j=i$.
 It means that the point $(i_a, j_a)$ in Fig. 2(a)/2(d) is the same with the point $(j_a, i_a)$
 in the excluded part of the figure.
 Figure 2(b)/2(c) displays the transfer
 of $l$-nucleotide strings from the coding/non-coding region to the non-coding/coding one.
 Only one half is kept due to the complementary symmetry about the diagonal line $j=i$.
 It means that the point $(j_a, i_a)$ in the excluded part of Fig. 2(b)
 is the same with the mirror symmetrical point $(j_a, i_a)$ of the point $(i_a, j_a)$ in the Fig. 2(c).
 So the point $(i_a, j_a)$ in Fig. 2(b) and the mirror symmetrical point $(j_a, i_a)$ in Fig. 2(c)
 can be combined into a figure, which displays the transfer of $l$-nucleotide strings
 from the coding region to the non-coding one.
 In the same way, a figure displaying
 the transfer of $l$-nucleotide strings from the non-coding region to the coding one
 can be generated by combining Fig. 2(c) and the mirror symmetrical part of Fig. 2(b).
 There exists the rotational symmetry about the diagonal line between two figures.
 It is evident that the transfer of $l$-nucleotide strings
 most and least frequently appears in the coding and non-coding regions, respectively.
 In Figs. 2(a) and 2(d), there appear some horizontal and vertical dense/sparse bands.
 In general, a vertical dense/sparse band
 $[i_{init}, i_{term}]$ means that $l$-nucleotide strings placed
 in the region $[i_{init}, i_{term}]$ of the genome are transferred to the more/less positions
 $[j, j+i_{term}-i_{init}]$ of the genome, where $j>i_{init}$. The more/less points
 $(i \in [i_{init}, i_{term}], j >i_{init})$  in the recurrence plot form the vertical dense/sparse band.
 A horizontal dense/sparse band $[j_{init},j_{term}]$ means that $l$-nucleotide strings
 in the more/less positions $[i-j_{term}+j_{init},i]$  of the genome are transferred
 into the region $[j_{init},j_{term}]$ of the genome, where $i<j_{term}$.
 The more/less points $(i<j_{term}, j \in [j_{init},j_{term}])$ in the recurrence
 plot form the horizontal dense/sparse band.
 Using the mirror symmetry, the horizontal dense/sparse bands
 in one figure can be reflected by the vertical ones in the excluded part
 of the same figure.
 In Figs. 2(b) and 2(c), there also exist some horizontal and vertical dense/sparse bands, respectively.
 In the same way, using the complementary symmetry, the horizontal dense/sparse bands
 in the excluded part of Fig. 2(b)
 can be reflected by the vertical dense/sparse ones in Fig. 2(c).
 The vertical dense/sparse bands in Fig. 2(c) imply that the
 $l$-nucleotide strings in several local positions of the non-coding region are transferred to
 more/less positions in the coding one.
  The horizontal dense/sparse bands in Fig. 2(b) imply that the $l$-nucleotide strings in
  more/less positions of the coding region are transferred into several local positions in the non-coding one.

 To depict the above transfer behaviors of the
 nucleotide strings in one or several vertical bands in Figs. 2(a)-2(d)
 the correlation intensity is calculated by using Eq. (3) and drawn in Fig. 3.
 It describes the length of transferred nucleotide strings with
 the same correlation distance in one band
 or the sum of lengths of them in several bands.
  In the whole range of transfer distance, many discrete values are distributed and
 denoted as non-periodic correlation structures.
 To depict correlation structures in local regions, the region $d \in [0,5000]$ is
 magnified in Fig. 3. Some equidistant parallel lines with a basic
 transferring length $d_{b_2} = 888$ appear in Fig. 3(a). Another basic
 transferring length $d_{b_1} = 6$ can be also determined when the region $d \in [0,200]$
  is further magnified in Fig. 3(a). They form periodic correlation structures.
 In the following subsections, we will analyze
   the periodic and non-periodic correlation structures.

\subsection{Correlation analysis of nucleotide strings}

\subsubsection{Periodic correlation structures}

By using the above method in Sect. 2, it is found that the periodic transfer of
nucleotide strings with lengths ($L \geq 20$) is confined in
local regions of the recurrence plot for $l$=15. Figures 4(a)-4(d) display several parallel
lines in the local regions, where basic transferring lengths are determined
as 6, 306, 18 and 888, respectively.
In general, parallel lines in a local region of the recurrence plot reflects periodic transfer
of several substrings in a long nucleotide string.
The periodic correlation structures are classified into three kinds
in terms of the relation of the substrings to the long nucleotide strings.

(1) The long nucleotide string is a periodic one
composed of one basic string, which is divided into several
substrings by gaps.
 The long nucleotide strings in Figs. 4(c) and 4(d) appear in the local
regions ($2082442-2082614$) and ($2354010-2358767$), respectively.
The local region in Fig. 4(c) is embedded in the coding region
 ($2080885-2082720$) of the gene $slr0422$.
 It is evident that there appear some
equidistant parallel lines with the basic transferring length $3d_{b1}=18$.
For each transfer distance, the total nucleotide string is divided into several
substrings with different lengths for transfer. The gap
between two neighboring substrings consists of one base only.
There exist 8 independent gaps, which are "t"
at 2082471, "c" at 2082480, "g" at 2082483, "c" at 2082498,
"c" at 2082534, "g" at 2082537, "c" at 2082588 and "c" at 2082606. The substrings
are transferred with integer times of the basic transferring length
to form periodic correlation structures as shown in Fig. 3(a).
 Once the 8 replaced bases "c", "t", "a", "t", "t", "a", "c" and "c" in the gaps
 are restored, respectively, the divided substrings
will combine to form a continuous periodic nucleotide string
and make the periodic transfer of the basic string.

Similarly, the local region in Fig. 4(d)
is embedded in the coding region
 ($2351323-2360412$) of the gene $slr0364$.
It is evident that there exist 6 equidistant parallel
lines, whose lengths decrease as the transfer distance increases.
For the transfer of nucleotide strings in the local
region, we reorder the transferred nucleotide strings in terms of
the increasing periods in Table I.
The transferred nucleotide strings have
the same starting position 2354010 and the basic
transferring length $d_{b2}$=888. For each transfer distance,
the total nucleotide string is divided into several
substrings with different lengths. The gap
between two neighboring substrings consists of one base only.
Since the gap may appear at either the original position 1 or transferred one 2,
we present all gaps and their restored bases at the position 1 or 2 in Table I.
It is found that there appear 5 independent gaps in the long
nucleotide string, which are denoted by square brackets in Table I.
Other gaps are just reappearance of them.
The 5 independent gaps are "a"
at 2355834, "t" at 2356965, "a" at 2359386, "t" at 2358741 in the first
transfer and "t" at 2357853 in the second transfer.
 Once the replaced bases "t", "c", "t", "c" and "c" in the gaps are restored,
 respectively, the divided substrings
will combine to form a continuous periodic nucleotide string
and make the periodic transfer of the basic string.
Different from Fig. 4(c), Fig. 4(d) also displays two groups of some short discrete
 parallel lines between two long ones with the distance $d_{b2}$ in
 the local region. It is found that two identical nucleotide strings with the length 294 are
 placed at the phases 3-296 and 594-887 in the basic string with the length $d_{b2}$.
 It means that the transfer of
 the nucleotide string in a period from the first position to the second one
 has the correlation distance 591. In the same way, the transfer of
 the nucleotide string between two periods from the second/first position to the first one
 has the correlation distance 297/888. The transfer of
 the nucleotide string between two periods from the first position to the first/second one
 has the correlation distance 888/(591+888) and so on so forth.
 Of course, the 5 gaps also divide the nucleotide string into several substrings
 for transfer to form the periodic correlation structures in Fig. 3(a).
 So, once the replaced bases in the gaps are restored, the divided
 substrings will combine and make the periodic transfer of the nucleotide
 string.

(2) The long nucleotide string is a periodic one
composed of two basic strings, which are divided into several
substrings by gaps.
 The long nucleotide string in Fig. 4(a) appears in the local
region ($527395-528016$), which is embedded in the coding region
 ($524346-529595$) of the gene $slr1753$.
 It is evident that there appear some
equidistant parallel lines with the basic transferring length $d_{b1}$=6.
Basically, the long nucleotide string is composed by two basic strings
with lengths 12 and 6. However, the second one disappears at some positions
in the long nucleotide string. There also exist 2 independent gaps with "t" at
527588 and "t" at 527594, which divides the long nucleotide string
 into several substrings. The substrings
are transferred with the integer times of the basic transferring length
to generate periodic correlation structures as shown in Fig. 3(a).
 Once the 2 replaced bases "c" and "g" in the gaps are restored, respectively,
 the divided nucleotide strings will combine to form a continuous periodic
 nucleotide string and make the periodic transfer of the basic strings.

(3) The long nucleotide string is not a periodic one but is composed of
several substrings for the transfer and non-transfer.
The long nucleotide string in Fig. 4(b) appears in the local
region ($589463-592418$), which is embedded in the coding region
 ($587228-593125$) of the gene $slr2046$.
It is evident that there appear some
equidistant parallel lines with the basic transferring length 306.
However, there does not exist the basic string with length 306 in the genome.
In the long nucleotide string, several substrings
can be transferred to generate the periodic coherence
structures, but others cannot.
So the transferred nucleotide strings cannot combine to form
a continuous periodic nucleotide string, but can be still transferred
periodically in the long nucleotide string.

\subsubsection{Non-periodic correlation structures}

Figure 5 displays local regions in the recurrence plot for the
non-periodic correlation structures, where exist several non-equidistant parallel lines.
 Consider the local region ($379000-379942$) denoted by zone 2 in Fig.
5(b), which covers the coding region
 ($379065-379913$) of the gene $slr1075$, as an example.
It is evident that the lengths of the non-equidistant
parallel lines are almost identical in the increasing of
transfer distance. For the non-periodic transfer of
nucleotide strings, we reorder the transferred nucleotide strings
in terms of the increasing of transfer distance in Table II.
For the first transfer distance 719258, the original nucleotide
string with length 943 located at
379000-379942 is divided into 12 substrings by 11 gaps and
transferred to the position 1098258-1099200 covering the coding region
(1098323-1099171) of the gene $slr1357$. The 12 nucleotide
substrings have different lengths. Each gap has just one base.
Since the change of base in the gap cannot be predicted to happen
at the original position 1 or the transferred one 2 in Table II,
the replacement of the base at the position 2 is taken as the gap
in the transfer process. And the gaps appearing first in the process
are defined as independent and are denoted by square brackets in Table II.
Other gaps in the following process are just reappearance of them.
Once the replaced bases in the gaps are restored, as given in the Table II,
 the continuous transfer of the original nucleotide
string would happen.
For the seventh, eighth and tenth transfer distances
2064934, 2155041 and 2718370, the transfer of the original
nucleotide string is similar to that for the first one in
covering the coding regions (2443999-2444847), (2534106-2534954)
and (3097436-3098284) of the genes $slr0352$, $slr0230$ and $slr0704$,
respectively. For the
second transfer distance 967132, the original nucleotide string
is divided into 16 substrings by 15 gaps and transferred to the
position 1346132-1347073 covering the coding region (1346197-1346556) of the gene $slr0856$
and the coding region (1346550-1347044) of the gene $slr0857$. Most of the gaps are just
the replacement of one base. In particular, the base "a" at the original
position 379367 is not replaced but just removed in the
transferred position 1346499 for the gap. Since the nucleotide string in the
position 2 has the global movement with one-base, the transfer distance
 is shrunk to 967131. For the third, fifth and sixth
transfer distances 1084390, 1247100 and 1856496, the
transfer of the original nucleotide string is similar to that
for the second one on covering three groups of the coding regions of two genes.
 They are the coding regions (1463455-1463814), (1463808-1464302), (1626165-1626524),
(1626518-1627012), (2235561-2235902) and (2235914-2236408) of the genes
$slr1715$, $slr1716$, $slr2112$, $slr2113$, $slr1936$ and $slr1937$, respectively.
For the fourth transfer distance 1237839,
the original nucleotide string with length 626 is divided into 8
substrings by 7 gaps and transferred to the position
1616839-1617473 covering the coding region (1616904-1617419) of the gene $slr1524$.
Most of the gaps are just the replacement of one
base. In particular, the bases "a" and "g" are
inserted in the positions 1617312 and 1617316 for the gaps, respectively, so the nucleotide
strings in the position 2 have global movements with one-base and two-bases.
The transfer distance is expanded into 1237840 and
1237841, respectively. For the ninth transfer distance 2716982,
the original nucleotide string with length 336 is divided into 4
substrings by 3 gaps and transferred to the position
3095982-3096322 covering the coding region (3096048-3096332) of the gene $ssr1175$.
Each gap is just the replacement of one base. There
also exist some short nucleotide strings with the similar
transfer to the above one, which are
not presented in Table II. Therefore, the non-periodic correlation
structure is generated by the transfer of non-periodic
nucleotide strings divided by several gaps.

In the $synecho$ genome, mobile elements with different lengths,
which referred to as selfish repetitive
DNA sequences, are provided. Due to the comparison
of the relative positions and lengths,
11 mobile elements located near the transferred nucleotide strings in the zone 2 are given
in Table II. It is evident that the mobile elements are almost identical
to the nucleotide strings at the transfer positions. The slight differences between them are due to the different choices of
their starting and ending positions.
From the above correlation analysis, the mobile
elements are almost the same or their substrings
and are transferred with the non-periodic distance.
In the transfer process, the mobile elements are divided into
the substrings by several gaps. Most of gaps are either the replacement of one
base or the insertion/reduction of one base.

The similar transfer of nucleotide strings with different lengths
for the non-periodic correlation structures appears in
other local regions. 6 independent cases
for the non-periodic transfer of the nucleotide strings with the lengths larger than 445 bases
are distributed in the local regions (52217-53173)  in Fig. 5(a),
(520876-521321) in Fig. 5(c), (573394-574580), (1200307-1201479) in Fig. 5(d),
(1483390-1484062) and (1614619-1615832),
which are denoted by zones 1, 3, 4, 5, 6 and 7, respectively.
On the one hand, the zones 1, 4, 5, and 6 cover the coding regions
(52260-53108), (573394-574580), (1200307-1201479) and (1483390-1484062)
  of the genes $sll1397$, $slr2036$, $sll1780$ and $ssr2898$, respectively.
The zone 3 covers the coding region (520918-521169) of the gene $ssl1922$
and overlaps with the coding region (521076-521357) of the gene $ssl1920$.
The zone 7 overlaps with the coding region (1614572-1615648) of the gene $slr522$.
On the other hand, the nucleotide strings in the zones are also transferred to
cover or overlap with coding regions of several genes in the genome.
In the same way, the nucleotide strings at the transfer positions in the 6 zones are
almost identical the mobile elements.

For the non-periodic transfer in each zone,
firstly, iterative transfer distance $x_k$
is determined by using Eq. (4). Two-dimensional reconstructed
vectors ${\bf y}_k$ generated by using Eq. (5) are then drawn in Fig. 6.
In the genome, the maximal
distance in the non-periodic transfer is about $1.4 \times 10^6$
 in the zones 4 and 7.
 Any transfer with a smaller distance disappears before it.
The second maximal distance in the non-periodic transfer
is about $9.5 \times 10^5$ in the zones 1, 2, 3 and 6.
It follows the transfer with a smaller
distance and then goes on or stops due to its position in the genome.
It is evident that some points in the reconstructed phase space
are situated along a line, demonstrating linear dependence.
By using the least square method, the fitting lines 1 and 2 drawn in Fig. 6 are determined as
$x_{k+1} = 954438 + 0.0874 (x_k - 323449)$ and
$x_{k+1} = 268827 + 13.47 (x_k -942503)$, respectively.
Line 1 reflects two steps for the continuous iterative transfer
until the second maximum. Line 2 describes the transfer for
a departure from the second maximum. The two fitting lines probably imply an
intrinsic dynamics in the transfer of nucleotide strings.

\section{Conclusion and discussions}

In summary, by using the recurrence plot method and the phase space reconstruction technique,
we have investigated the
transfer properties of nucleotide strings in the $synecho$ genome
and demonstrated the presence of periodic and non-periodic correlation structures.
The periodic correlation structures are generated by periodic transfer of
several substrings in long periodic or non-periodic nucleotide strings
embedded in the coding regions of genes.

The non-periodic correlation structures are generated by non-periodic transfer
of several substrings covering or overlapping with the coding regions of genes.
In the periodic and non-periodic transfer,
some gaps divide the long nucleotide strings into the substrings
and prevent their global transfer.
Most of the gaps are either the replacement of one base
or the insertion/reduction of one base.
 In the reconstructed phase
space, the points generated from two or three steps for the continuous iterative
transfer via the second maximal distance can be fitted by two lines.
It partly reveals an
intrinsic dynamics in the transfer of nucleotide strings.
Due to the comparison of the relative positions and lengths,
the substrings concerned with the non-periodic correlation structures are almost identical to
the mobile elements annotated in the genome. The mobile elements have been thus endowed
with the basic results on the correlation structures.

Although the repeats of nucleotide strings in the genome may be determined by the general $k$-mer method,
the correlation analysis can reflect the relative positions among the repeated nucleotide strings in the genome
and their internal structures generated by gaps.
 The periodic and non-periodic
structures in the cording and non-cording regions of the genome revealed by the correlation analysis may
relate to the heredity and variance of the cells:
the transfer of continuous/interrupted nucleotide strings
in the genome keeps/changes the nucleotide composition for the heredity/variance.
Moreover, the junk DNA of a genome includes of many transferable elements
in non-cording regions. Its unknown biological functions in cells are covered by
the mystery of transfer of the elements in the whole genome.
The proposed periodic and non-periodic correlation structures may have fundamental
importance for the biological functions of the junk DNA.

\newpage
\textbf{Acknowledgments} We would like to thank the National
Science Foundation for partial support through the Grant No.
11172310 and the IMECH/SCCAS SHENTENG 1800/7000 research computing
facilities for assisting in the computation.

\newpage
{\footnotesize Table I. Periodic transfer of nucleotide
strings with lengths $L(\geq 20)$ in the local region
($2.354-2.36 \times 10^6$) for the $synecho$ genome.
$d_T$($d_{b2}$) is the (basic) transfer distance.
$L_T$ is the total lengths of transferred nucleotide
strings. $s_1 \rightleftharpoons s_2$ denotes the restoration
of bases in gaps.

\begin{tabular}{llllll}
\hline
$k$ &$d_T$($L_T$) & Position 1 & $L$ & Position 2 & $s_1 \rightleftharpoons s_2$\\
\hline

1 &$d_{b2}$(4753) & 2354010 - 2354945  & 936  & 2354898 - 2355833 & $t \rightarrow [a]$\\
  &  & 2354947 - 2355833  & 887 & 2355835 - 2356721 & $a \leftarrow t$\\
  &  & 2555835 - 2356076  & 242 & 2356723 - 2356964 & $c \rightarrow [t]$\\
  &  & 2356078 - 2358497  & 2420& 2356966 - 2359385 & $t \rightarrow [a]$\\
  &  & 2358499 - 2358740  & 242 & 2359387 - 2359628 & $[t] \leftarrow c$\\
  &  & 2358742 - 2358767  & 26  & 2359639 - 2359655\\
\hline

2 &2$d_{b2}$(3864) & 2354010  - 2354057 & 48   & 2355786 - 2355833 & $t \rightarrow a$\\
  &           & 2354059  - 2355188 & 1130 & 2355835 - 2356964 & $c \rightarrow t$\\
  &           & 2355190  - 2355833 & 644  & 2356966 - 2357609 & $a \leftarrow t$\\
  &           & 2355835  - 2356076 & 242  & 2357611 - 2357852 & $c \rightarrow [t]$\\
  &           & 2356078  - 2357609 & 1532 & 2357854 - 2359385 & $t \rightarrow a$\\
  &           & 2357611  - 2357852 & 242  & 2359387 - 2359628 & $t \leftarrow c$\\
  &           & 2357854  - 2357879 & 26   & 2359630 - 2359655 & \\
\hline

3 &3$d_{b2}$(2976) & 2354010  - 2354300 & 291  & 2356674 - 2356964 & $c \rightarrow t$\\
  &           & 2354302  - 2355188 & 887  & 2356966 - 2357852 & $c \rightarrow t$\\
  &           & 2355190  - 2355833 & 644  & 2357854 - 2358497 & $a \leftarrow t$\\
  &           & 2355835  - 2356076 & 242  & 2358499 - 2358740 & $c \rightarrow t$\\
  &           & 2356078  - 2356721 & 644  & 2358742 - 2359385 & $t \rightarrow a$\\
  &           & 2356723  - 2356964 & 242  & 2359387 - 2359628 & $t \leftarrow c$\\
  &           & 2356966  - 2356991 & 26   & 2359630 - 2359655 & \\
\hline

4 &4$d_{b2}$(2092) & 2354010  - 2354300 & 291  & 2357562 - 2357852 & $c \rightarrow t$\\
  &           & 2354302  - 2355188 & 887  & 2357854 - 2358740 & $c \rightarrow t$\\
  &           & 2355290  - 2356103 & 914  & 2358742 - 2359655 & \\
\hline

5 &5$d_{b2}$(1204) & 2354010  - 2354300 & 291  & 2358450 - 2358740 & $c \rightarrow t$\\
  &           & 2354302  - 2354945 & 644  & 2358742 - 2359385 & $t \rightarrow a$\\
  &           & 2354947  - 2355215 & 269  & 2359387 - 2359655 & \\
\hline

6 &6$d_{b2}$(317) & 2354010  - 2354057 & 48   & 2359338 - 2359385 &$t \rightarrow a$\\
  &           & 2354059  - 2354327 & 269  & 2359387 - 2359655 &\\
\hline
\end{tabular}
 }

\newpage
{\footnotesize Table II. Non-periodic transfer of nucleotide
strings in the local region ($3.79-3.8 \times 10^5$) for the
$synecho$ genome. Notation as in Table I. The mobile elements (with lengths)
located near the transferred nucleotide strings are given in the last line
of each transfer step.

\begin{tabular}{llllll}
\hline
$k$ & $d_T$($L_T$) & Position 1 & $L$  & Position 2 & $s_1 \rightarrow s_2$\\
\hline

1 &719258(932) & 379000 - 379037  & 38 & 1098258 - 1098295 & $g \rightarrow [a]$\\
  &  & 379039  - 379084   & 46  & 1098297 - 1098342 & $a \rightarrow [g]$\\
  &  & 379086  - 379093   & 8   & 1098344 - 1098351 & $g \rightarrow [a]$\\
  &  & 379095  - 379435   & 341 & 1098353 - 1098693 & $c \rightarrow [t]$\\
  &  & 379437  - 379615   & 179 & 1098695 - 1098873 & $a \rightarrow [g]$\\
  &  & 379817  - 379630   & 14  & 1098875 - 1098888 & $t \rightarrow [c]$\\
  &  & 379632  - 379669   & 38  & 1098890 - 1098927 & $a \rightarrow [t]$\\
  &  & 379671  - 379678   & 8   & 1098929 - 1098936 & $a \rightarrow [g]$\\
  &  & 379680  - 379682   & 3   & 1098938 - 1098940 & $c \rightarrow [t]$\\
  &  & 379684  - 379922   & 239 & 1098942 - 1099180 & $c \rightarrow [t]$\\
  &  & 379924  - 379939   & 16  & 1099182 - 1099197 & $t \rightarrow [c]$\\
  &  & 379941  - 379942   & 2   & 1099199 - 1099200 & \\
  & ME(947)& 378993 - 379939   &     & 1098251 -1099197 & \\
\hline

2 &967132(361) & 379000 - 379007 & 8 & 1346132 - 1346140 & $a \rightarrow [g]$\\
  &  & 379009 - 379032  & 24  & 1346141 - 1346164 &$t \rightarrow [c]$\\
  &  & 379034 - 379046  & 13  & 1346166 - 1346178 &$a \rightarrow [g]$\\
  &  & 379048  - 379084   & 37  & 1346180 - 1346216 &$a \rightarrow g$\\
  &  & 379086  - 379093   & 8  & 1346218 - 1346225 &$g \rightarrow a$\\
  &  & 379095  - 379161   & 67  & 1346227 - 1346293 &$c \rightarrow [t]$\\
  &  & 379163  - 379366   & 204 & 1346295 - 1346498 &$a \rightarrow [\ ]$\\
  &967131(567) & 379368  - 379373   & 6   & 1346499 - 1346504 &$c \rightarrow [t]$\\
  &  & 379375 - 379594  & 220  & 1346506 - 1346725 &$a \rightarrow [g]$\\
  &  & 379596  - 379792   & 197  & 1346727  - 1346923 & $g \rightarrow [a]$\\
  &  & 379594  - 379795   & 2  & 1346925  - 1346926 & $a \rightarrow [g]$\\
  &  & 379797  - 379894   & 98   & 1346928  - 1347025 &$t \rightarrow [c]$\\
  &  & 379896  - 379914   & 19   & 1347027  - 1347045 &$g \rightarrow [a]$\\
  &  & 379916  - 379921   & 6    & 1347047  - 1347052 &$t \rightarrow [c]$\\
  &  & 379923  - 379932   & 10   & 1347054  - 1347063 &$a \rightarrow [t]$\\
  &  & 379934  - 379942   & 9    & 1347065  - 1347073 &\\
  &ME(946)  &                    &      & 1346125  - 1347070 &\\

\hline
\end{tabular}

\begin{tabular}{lllllll}
\hline
3 &1084390(367)      & 379000  - 379037  & 38    & 1463390 - 1463427 & $g \rightarrow a$\\
  &          & 379039  - 379366  & 328   & 1463429 - 1463756 & $a \rightarrow [\ ]$\\
  &1084389(579)      & 379368  - 379435  & 68    & 1463757 - 1463824 & $c \rightarrow t$\\
  &          & 379437  - 379615  & 179   & 1463826 - 1464004 & $a \rightarrow g$\\
  &          & 379617  - 379630  & 14  & 1464006 - 1464019 & $t \rightarrow c$\\
  &          & 379632  - 379678  & 47    & 1464021 - 1464067 & $a \rightarrow g$\\
  &          & 379680  - 379922  & 243    & 1464069 - 1464311 & $c \rightarrow t$\\
  &          & 379924  - 379942  & 19     & 1464313 - 1464331 & \\
  & ME(946)         &                   &        & 1463383 - 1464328 & \\
\hline

4 &1237839(470)      & 379000  - 379037  & 38      & 1616839 - 1616876 & $g \rightarrow a$\\
  &           & 379039  - 379435  & 397    & 1616878 - 1617274 & $c \rightarrow t$\\
  &           & 379437  - 379469  & 33     & 1617276 - 1617308 & $g \rightarrow [a]$\\
  &           & 379471  - 379472  & 2     & 1617310 - 1617311 & $g \rightarrow [ac]$\\
  &1237840(2)   & 379474  - 379475  & 2     & 1617314 - 1617315 & $t \rightarrow [gc]$\\
  &1237841(154) & 379477  - 379615  & 139      & 1617318 - 1617456 & $a \rightarrow g$\\
  &           & 379617  - 379625  & 9      & 1617458 - 1617466 & $t \rightarrow [c]$\\
  &           & 379627  - 379632  & 6      & 1617468 - 1617473 & \\
  & ME(678)          &                   &        & 1616832 - 1617509 & \\
\hline

5 &1247100(365)      & 379000   - 379037  & 38     & 1626100 - 1626137 & $g \rightarrow a$\\
  &           & 379039   - 379293  & 255    & 1626139 - 1626393 & $c \rightarrow [t]$\\
  &           & 379295   - 379366  & 72      & 1626395 - 1626466 & $a \rightarrow $\\
  &1247099(569) & 379368   - 379435  & 68     &  1626467 - 1626534 & $c \rightarrow t$\\
  &           & 379437   - 379615  & 179    & 1626536 - 1626714 & $a \rightarrow g$\\
  &           & 379417   - 379630  & 14     & 1626716 - 1626729 & $t \rightarrow c$\\
  &           & 379632   - 379678  & 47    & 1626731 - 1626777 & $a \rightarrow g$\\
  &           & 379680   - 379922  & 243    & 1626779 - 1627021 & $a \rightarrow t$ \\
  &           & 379624   - 379942  & 19      & 1627023 - 1627041 &  \\
  &  ME(946)         &                    &         & 1626093 - 1627038 &  \\
\hline

6 &1856496(366) & 379000   - 379037  & 38    & 2235496 - 2235533 & $g \rightarrow a$\\
  &           & 379039   - 379366  & 328    & 2235535 - 2235862 & $a \rightarrow  $\\
  &1856495(569) & 379368   - 379435  & 68     & 2235863 - 2235930 & $c \rightarrow t$\\
  &           & 379437   - 379615  & 179    & 2235932 - 2236110 & $a \rightarrow g$\\
  &           & 379617   - 379630  & 14     & 2236112 - 2236125 & $t \rightarrow c$\\
  &           & 379632   - 379678  & 47     & 2236127 - 2236173 & $a \rightarrow g$\\
  &           & 379680   - 379922  & 243    & 2236175 - 2236417 & $c \rightarrow t$\\
  &           & 379924   - 379941  & 18     & 2236419 - 2236436 & $t \rightarrow a$\\
  & ME (946)         &                    &        & 2235496 - 2236436 &\\
\hline
\end{tabular}

\begin{tabular}{llllllll}
\hline
7 &2064934(936) & 379000   - 379037  & 38     & 2443934 - 2443971 & $g \rightarrow a$\\
  &           & 379039   - 379435  & 397    & 2443973 - 2444369 & $c \rightarrow t$\\
  &           & 379437   - 379539  & 103    & 2444371 - 2444473 & $g \rightarrow [a]$\\
  &           & 379541   - 379615  & 75     & 2444475 - 2444549 & $a \rightarrow g$\\
  &           & 379517   - 379630  & 14     & 2444551 - 2444564 & $t \rightarrow c$\\
  &           & 379632   - 379678  & 47     & 2444566 - 2444612 & $g \rightarrow a$\\
  &           & 379680   - 379922  & 243    & 2444614 - 2444856 & $c \rightarrow t$\\
  &           & 379924   - 379942  & 19     & 2444858 - 2444876 & \\
  & ME (947)          &                    &        & 2443927 - 2444873 & \\
\hline

8 &2155041(935)      & 379000   - 379037  & 38      & 2534041 - 2534078 & $g \rightarrow a$\\
  &           & 379039   - 379203  & 165    & 2534080 - 2534244 & $c \rightarrow [t]$\\
  &           & 379205   - 379435  & 231    & 2534246 - 2534476 & $c \rightarrow t$\\
  &           & 379437   - 379615  & 179    & 2534478 - 2534656 & $g \rightarrow a$\\
  &           & 379417   - 379630  & 14     & 2534458 - 2534671 & $c \rightarrow t$\\
  &           & 379632   - 379678  & 47     & 2534673 - 2534719 & $g \rightarrow a$\\
  &           & 379680   - 379922  & 243    & 2534721 - 2534963 & $c \rightarrow t$\\
  &           & 379924   - 379941  & 18     & 2534965 - 2534982 & $t \rightarrow [c]$\\
  &  ME (947)        &                    &        & 2534034 - 2534980 &\\
\hline

9 &2716982(338)      & 379000   - 379037  & 38     & 3095982 -3096019 & $g \rightarrow a$\\
  &           & 379039   - 379203  & 165    & 3096201 -3096185 & $c \rightarrow t$\\
  &           & 379205   - 379337  & 133     & 3096187 -3096319 & $g \rightarrow [a]$\\
  &           & 379339   - 379340  & 2      & 3096121 -3096322 &\\
  &  ME (345)        &                    &        & 3095975 -3096319 &\\
\hline

10 &2718370(936)      & 379000   - 379037  & 38     & 3097370 -3097407 & $g \rightarrow a$\\
   &          & 379039   - 379203  & 165    & 3097409 -3097573 & $c \rightarrow t$\\
   &          & 379205   - 379435  & 231    & 3097575 -3097805 & $c \rightarrow t$\\
   &          & 379437   - 379615  & 179   & 3097807 -3097985 & $a \rightarrow g$\\
   &          & 379617   - 379630  & 14     & 3097987 -3098000 & $c \rightarrow t$\\
   &          & 379632   - 379678  & 47     & 3098002 -3098048 & $a \rightarrow g$\\
   &          & 379680   - 379922  & 243    & 3098050 -3098292 & $c \rightarrow t$\\
   &          & 379924   - 379942  & 19      & 3098294 -3098312&\\
   &  ME (947)       &                    &        & 3097363 -3098309 &\\
\hline
\end{tabular}
}

\newpage
\textbf{Figure caption}

Fig.~1. Mutual information function $I(m)$ in the log scale for the $synecho$ genome. A local
blow-up region $m\in[0,100]$ with the linear scale is redrawn in the figure.

Fig.~2. Recurrence plots (a) in the coding region; (b) from
the coding region to non-coding one; (c) from non-coding
region to coding one; (d) in the non-coding region of
the $synecho$ genome for 15-nucleotide strings.

Fig.~3. Correlation intensity $\Xi(d)$ plots versus
transfer distance $d$
(a) in the coding region; (b) from
the coding region to non-coding one; (c) from non-coding
region to coding one; (d) in the non-coding region of
the {\it synecho} genome. Two local blow-up regions near $d=0$ are redrawn in Fig. 3(a).

Fig.~4. Four local regions in the recurrence plot for the {\it
synecho} genome, where periodic correlation structures are exhibited.
The diagonal line ($i=j$) is also plotted.

Fig.~5. Four local regions (52217-53173), (379000-379942), (520876-521321) and (1200307-1201479)
denoted by zones 1, 2, 3 and 5 in the recurrence plot for the {\it
synecho} genome, where non-periodic correlation
structures are exhibited. The diagonal line ($i=j$) is also plotted.

Fig.~6. Two-dimensional vectors in the phase space reconstructed from
the iterative transfer distances in the 7 zones.

\end{document}